\documentclass[journal,10pt]{IEEEtran}
\usepackage{algorithm} 
\usepackage{algorithmic} 
\usepackage{multirow} 
\usepackage{amsmath}
\usepackage{xcolor}
\usepackage{amsmath}
\usepackage{stfloats}
\usepackage{amssymb}
\usepackage{amsmath}
\usepackage{cite}
\usepackage{url}
\usepackage{xcolor}
\usepackage{cite,graphicx,amsmath,amssymb}
\usepackage{subfigure}
\usepackage{citesort}
\usepackage{fancyhdr}
\usepackage{mdwmath}
\usepackage{mdwtab}
\usepackage{caption}
\usepackage{amsthm}
\usepackage{enumitem}
\usepackage{setspace}
\setenumerate{fullwidth,itemindent=\parindent,listparindent=\parindent,itemsep=0ex,partopsep=0pt,parsep=0ex}

\newtheorem{theorem}{Theorem}

\newtheorem{lemma}{Lemma}

\newtheorem{Definition}{Definition}

\newtheorem{corollary}{Corollary}

\makeatletter
\def\ScaleIfNeeded{%
\ifdim\Gin@nat@width>\linewidth \linewidth \else \Gin@nat@width
\fi } \makeatother

\makeatletter
\renewcommand{\maketag@@@}[1]{\hbox{\m@th\normalsize\normalfont#1}}%
\makeatother

\begin{document}

\title{Hybrid NOMA for STAR-RIS Enhanced Communication}

\author{
        Jiayi~Lei,
        Tiankui~Zhang,~\IEEEmembership{Senior Member,~IEEE,}
        Yuanwei~Liu,~\IEEEmembership{Senior Member,~IEEE}
\vspace{-9pt}

\thanks{ Jiayi~Lei and Tiankui Zhang are with the School of Information and Communication Engineering, Beijing University of Posts and Telecommunications, Beijing 100876, China (e-mail: \{leijiayi,zhangtiankui \}@bupt.edu.cn).}
\thanks{Yuanwei Liu is with the School of Electronic Engineering and Computer Science, Queen Mary University of London, London E1 4NS, U.K. (e-mail: yuanwei.liu@qmul.ac.uk).}
}

\maketitle

\begin{abstract}
In this paper, a hybrid non-orthogonal multiple access (NOMA) framework for the simultaneous transmitting and reflecting reconfigurable intelligent surface (STAR-RIS) enhanced cell-edge communication is investigated. Specifically, one transmitted user and one reflected user are paired as one NOMA-pair, while multiple NOMA-pairs are served via time division multiple access (TDMA). The objective is to maximize the minimum downlink rate by jointly optimizing the user pairing, decoding order, passive beamforming, power and time allocation. A novel two-layer iterative algorithm is proposed to solve the highly coupled problem. Simulation results show that: 1) the proposed framework outperforms the conventional reflecting-only-RIS-based and the OMA-based frameworks; 2) the beamforming design and power allocation dominate the achieved performance; 3) increasing the number of passive elements and shortening the distance between BS and STAR-RIS are two effective ways to further improve the performance.

\end{abstract}
\providecommand{\keywords}[1]{\textbf{\textit{Index terms---}}#1}


\section{Introduction}
In traditional wireless communication networks, cell-edge users always suffer from poor quality-of-service (QoS) due to severe channel fading and inter-cell interference. For such practical scenarios, the reconfigurable intelligent surface (RIS) is a promising technology to enhance communications for cell-edge users. With the ability to smartly reconfigure the wireless propagation environment, the RIS is able to provide additional and high-quality transmission links \cite{RIS_magzine_zhangrui}. In addition, thanks to its small size, light weight, and high extensibility, the RIS is easy to be deployed in existing networks. However, the conventional RIS is designed to only reflect the incident signals, making the source and the destination have to be on the same side of the RIS. To address this issue, a novel simultaneous transmitting and reflecting RIS (STAR-RIS) is proposed in \cite{STAR_magazine_liuyuanwei}. The incident signal can be transmitted and reflected by the STAR-RIS simultaneously, and thus a full-space smart radio environment is created. With apparent advantages, the STAR-RIS aided networks have attracted widespread attention \cite{STAR1,STAR3}. In \cite{STAR1}, the coverage probability of a STAR-RIS assisted massive multiple-input multiple-output (MIMO) system was analyzed with Rayleigh fading and phase-shift errors. In \cite{STAR3}, the transmit power for a STAR-RIS aided multiple-input single-output system was minimized by joint active and passive beamforming design.

Non-orthogonal multiple access (NOMA) is another promising technology for future wireless communications.
It has been proved that NOMA yields a significant gain over conventional orthogonal multiple access techniques in terms of spectral efficiency and user fairness \cite{NOMA_magzine}.
Significantly, combining NOMA and STAR-RIS is a meaningful research topic. On the one hand, the STAR-RIS is able to reconfigure channels smartly and introduce additional degrees-of-freedom (DoFs) for system design, thus enhancing the NOMA gain.
On the other hand, due to the simultaneous transmission and reflection characteristics of STAR-RIS, users are naturally divided into transmission users and reflection users, which provides a basis for user pairing in NOMA.
Inspired by this, there emerge some works which combine the two techniques \cite{STAR+NOMA01,STAR_NOMA5,STAR+NOMA03}. In \cite{STAR+NOMA01}, the effective capacity of a STAR-RIS assisted NOMA network with two users on different sides of the STAR-RIS was studied. In \cite{STAR_NOMA5}, the authors investigated the secrecy performance of the STAR-RIS assisted NOMA networks. A sum rate maximization problem for STAR-RIS-NOMA systems was investigated in \cite{STAR+NOMA03}, where the decoding order, power allocation, active beamforming, and passive beamforming were jointly optimized.

Although some excellent works on STAR-RIS assisted NOMA networks have been conducted, most of them considered pure NOMA, which will face the following bottleneck. With a large number of users, it is necessary to perform user clustering to ensure the effectiveness of NOMA. In this case, users served by pure NOMA suffer from not only intra-cluster interference but also inter-cluster interference, which makes a severely negative effect on QoS and increases the complexity of network design.
Hybrid NOMA is one of the effective solutions to tackle this issue, but it has not yet received widespread attention \cite{RIS_hybrid1,RIS_hybrid2,STAR+FDMA_NOMA}. In \cite{RIS_hybrid1}, the authors studied hybrid NOMA and time division multiple access (TDMA) for the uplink transmission in reflecting-only RIS assisted wireless powered communication networks. Similarly, a hybrid TDMA-NOMA scheme was developed to balance the performance and signalling overhead for RIS aided Internet-of-Things (IoT) system in \cite{RIS_hybrid2}. However, as the functional limitation of reflecting-only RIS, only users located on one side of the RIS were considered in \cite{RIS_hybrid1,RIS_hybrid2}.
Furthermore, few related works investigate hybrid NOMA for STAR-RIS aided networks \cite{STAR+FDMA_NOMA}. A frequency division multiple access (FDMA)-NOMA mixed framework was applied in \cite{STAR+FDMA_NOMA} to eliminate inter-cluster interference. However, it is worth noting that the passive beamforming at STAR-RIS is time-selective, but not frequency-selective. As a result, the advantages of the STAR-RIS can not be fully utilized.

Motivated by the above observation, a novel hybrid NOMA framework for STAR-RIS assisted cell-edge networks is proposed in this paper, which reduces the complexity of network design while fully leveraging the advantages of STAR-RISs. The main contributions are summarized as follows: 1) We propose a novel hybrid NOMA-TDMA framework for STAR-RIS assisted cell-edge networks, where one transmitted user and one reflected user are paired and served via NOMA, while multiple NOMA-pairs are served by TDMA; 2) We formulate a max-min rate problem and solve it by a novel two-layer iterative algorithm, where user pairing, passive beamforming, power and time allocation are jointly optimized; 3) Numerical results show the superiority of the proposed network framework and the two-layer algorithm. Moreover, increasing the number of STAR-RIS elements and shortening the distance from STAR-RIS to BS are confirmed to be two effective ways to improve network performance.

\emph{Notation}: Scalars, vectors and matrices are denoted by Italic letters, bold-face lower-case, and bold-face upper-case, respectively. For a complex-valued vector $\mathbf{a}$, $\mathbf{a}^{\mathrm{T}}$ means its transpose, $\mathbf{a}^{\mathrm{H}}$ means its conjugate transpose, and $diag(\mathbf{a})$ denotes a diagonal matrix with the elements of vector $\mathbf{a}$ on the main diagonal. Besides, $\left\| {\cdot} \right\|$ denotes a vector's Euclidean norm, and $arg(\cdot)$ denotes a complex number's argument.

\vspace{-0.1cm}
\section{System Model and Problem Formulation}

\subsection{System Model}
\begin{figure}[htb]
  \centering
  \includegraphics[width=3in]{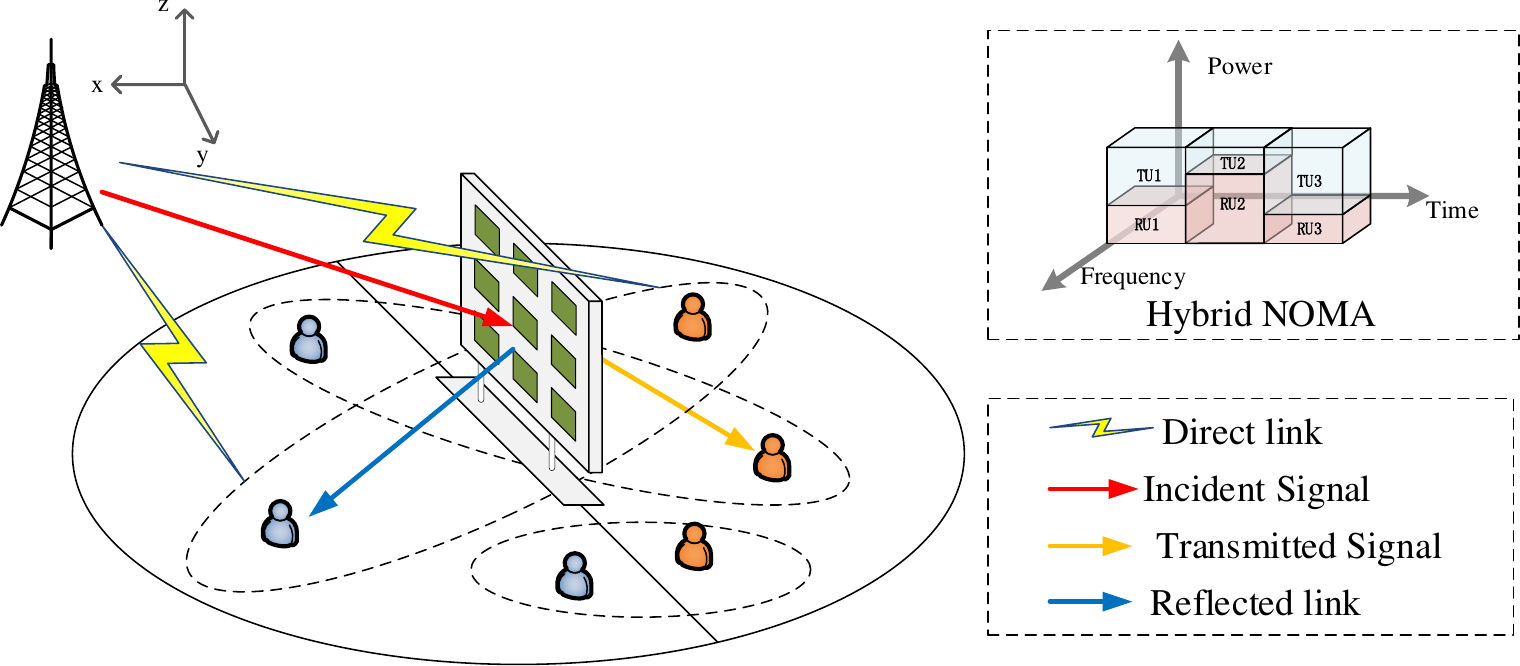}\\
  \caption{STAR-RIS assisted communication networks.}\label{Fig1}
\end{figure}

As shown in Fig. \ref{Fig1}, we consider a cell-edge area, where multiple users are randomly distributed, and a STAR-RIS is deployed in the center to enhance the communications. For the convenience of explanation, a 3D Cartesian coordinate system is established. Let ${\mathbf{q}_S} = [{x_S},{y_S},{z_S}]$, ${\mathbf{q}_B} = [{x_B},{y_B},{z_B}]$ and ${\mathbf{q}_k} = [{x_k},{y_k},0]$  denote the 3D positions of the STAR-RIS first element, the base station (BS) and cell-edge users respectively. The BS and users are all equipped with a single antenna. The STAR-RIS consists of a uniform planar array (UPA) with $M = {M_y}{M_z}$ passive transmitting and reflecting elements, where $M_y$ and $M_z$ denote the number of elements along the $y-$ and $z-$axis, respectively. The STAR-RIS adopts the energy-splitting protocol, i.e., each element of the STAR-RIS splits the incident signal into transmitted and reflected signals to serve users at both sides of the surface simultaneously  \footnote{There are three practical protocols for operating STAR-RISs: namely energy splitting (ES), mode switching (MS), and time switching (TS). Based on the insights obtained in \cite{STAR_protocols}, the ES protocol is the best option among the three protocols for commutations with high quality-of-service (QoS) requirements. Therefore, we adopt the ES protocol in this paper. Numerical simulation and comparison of the three protocols in STAR-RIS assisted wireless communications will be considered in our future study.}. As the whole region is divided into the right-half sub-region and the left-half sub-region by the STAR-RIS, users can be divided into transmitted users (TU) and reflected users (RU), whose sets are denoted as ${\mathcal{K}_T}$ and ${\mathcal{K}_R}$ respectively. We assume that users are equally located in the two sub-regions, that is, $|{\mathcal{K}_T}|= |{\mathcal{K}_R}| = K$ \footnote{ The proposed network framework and algorithm in this paper can be extended to cases where the number of TUs is not equal to that of RUs, with small modifications on the design of user pairing.}.

We consider that one TU is paired with one RU, and then all users are grouped into $K$ user-pairs. Let binary variable ${c_{{k_T},{k_R}}}({c_{{k_R},{k_T}}}) \in \{ 0,1\} ,{k_T} \in {{\cal K}_T},{k_R} \in {{\cal K}_R}$ denote whether the TU $k_T$ and the RU $k_R$ are paired up, which satisfies $\sum\limits_{{k_R} = 1}^K {{c_{{k_T},{k_R}}}}  = 1$. The two paired users receive data from the BS via non-orthogonal multiple access (NOMA). $K$ user-pairs are served sequentially via TDMA. This multiple access scheme is referred to as hybrid NOMA in this paper. Given $k \in {{\cal K}_T} \cup {{\cal K}_R}$, let $\bar{k}$ denote the index of the user which is paired with ${k}$, i.e., $c_{{k},\overline {k}} = 1$. Then, the time allocation coefficient of the user-pair $\{ {k},
\bar{k} \} $ is denoted by ${\tau _{\{ {k},
\bar{k} \} }} \in [0,1]$, which is supposed to satisfy \begin{small}$\sum\limits_{k \in {{\cal K}_T}({k \in {{\cal K}_R}})} {{\tau _{\{ k,\bar{k} \} }} = 1}$\end{small}.

As user-pairs occupy orthogonal time resource in the hybrid NOMA framework, the passive beamforming vectors of the STAR-RIS can be designed specifically and independently for each user-pair, so as to reconstruct the channel between the BS and the currently served user-pair more purposefully. Let \begin{small}${\mathbf{V}}_{k} = \sqrt {\beta_k}diag\left( {{e^{j\theta _k^1}}, {e^{j\theta _k^2}},...,{e^{j\theta _k^M}}} \right)$ \end{small} denote the transmission and reflection coefficient matrix of STAR-RIS for the user $k$, where $\sqrt {{\beta _k}} \in [0,1]$ denotes the amplitude and $\theta _k^m$ denotes the phase shifts of $m$-th STAR-RIS element. We assume that there is no energy dissipated by STAR-RIS, that is, $\beta _k + \beta _{\bar{k}} = 1$.

\subsection{Channel Model}
There are two kinds of downlinks for each user, direct link and STAR-RIS assisted link. Let ${h_{B,k}} \in {\mathbb{C}^{1 \times 1}}$, ${\mathbf{h}_{B,S}} \in {\mathbb{C}^{M \times 1}}$ and ${\mathbf{h}_{S,k}} \in {\mathbf{}\mathbb{C}^{M \times 1}}$ denote the channel vectors from the BS to user, from the BS to the STAR-RIS, and from the STAR-RIS to user respectively. Considering the blocked line-of-sight (LoS) link and potential extensive scattering from the BS to users, the propagation ${h_{B,k}} \in {\mathbb{C}^{1 \times 1}}$ is modeled as Rayleigh fading, while the transmission links from the BS to the STAR-RIS and from the STAR-RIS to users are modeled as LoS-dominated channels.
Specifically,  the channel coefficient of the direct link is expressed as
\begin{small}
\begin{equation}
 \vspace{-0.1cm}
{h_{B,k}} = \sqrt {\frac{{{\delta _0}}}{{{d_{B,k}}^{{\alpha _1}}}}} {\widetilde h_{B,k}}.
 \vspace{-0.1cm}
\end{equation}
\end{small}where $\delta _0$ is the channel power at the reference distance of $1$m, $\alpha_1$ is the path loss exponent, ${d_{B,k}} = \left\| {{{\mathbf{q}}_B} - {{\mathbf{q}}_k}} \right\|$ is the distance between the BS and the user $k$, and $\widetilde h_{B,k}$ is complex Gaussian random variable with zero mean and unit variance, i.e., $\widetilde h_{B,k}\sim\mathcal{CN}(0,1)$.

The channel vectors from the BS to the STAR-RIS and from the STAR-RIS to users are given by
\begin{equation}
{\mathbf{h}_{B,S}} = \sqrt {\frac{{{\delta_0}}}{{{d_{B,S}}^{\alpha _2} }}}{\mathbf{a}_{AoA}},
\end{equation}
\begin{equation}
\begin{array}{l}
{\mathbf{h}_{S,k}} = \sqrt {\frac{{{\delta_0}}}{{{d_{S,k}}^{\alpha _3} }}}{\mathbf{a}_{AoD}},
\end{array}
\end{equation}
where ${d_{B,S}} = \left\| {{{\mathbf{q}}_B} - {{\mathbf{q}}_S}} \right\|$ and ${d_{S,k}} = \left\| {{\mathbf{q}}_S} - {{\mathbf{q}}_k} \right\|$. ${\mathbf{a}_{AoA}}$ is the receive array response vector and  given as \begin{scriptsize} ${{\mathbf{a}_{AoA}}} = {e^{ - j\frac{{2\pi {d_{U,S}}}}{\lambda }}}{{\left[ {1,{e^{ - j\frac{{2\pi {d_{y}}}}{\lambda }\sin {\varphi _r}\cos {\eta _r}}}...,{e^{ - j\frac{{2\left({{M_y} - 1} \right)\pi {d_y}}}{\lambda }\sin {\varphi _r}\cos {\eta _r}}}} \right]}^\mathrm{T}}\otimes {{{\left[ {1,{e^{ - j\frac{{2\pi {d_z}}}{\lambda }\sin {\varphi _r}\sin {\eta _r}}}...,{e^{ - j\frac{{2\left( {{M_z} - 1} \right)\pi {d_z}}}{\lambda }\sin {\varphi _r}\sin {\eta _r}}}} \right]}^\mathrm{T}}}$ \end{scriptsize}\hspace{-0.5em}, \hspace{-0.5em} where $\varphi _r$ and $\eta _r$ are the zenith angle of arrival (AoA) and the azimuth AoA of the signal from the BS to the STAR-RIS respectively. ${\mathbf{a}_{AoD}}$ is the transmit array response vector and given as \begin{scriptsize} ${\mathbf{a}_{AoD}} = {e^{ - j\frac{{2\pi {d_{S,k}}}}{\lambda }}}{\left[ {1,{e^{ - j\frac{{2\pi {d_y}}}{\lambda }\sin {\varphi _t}\cos {\eta _t}}}...,{e^{ - j\frac{{2\left( {{M_y} - 1} \right)\pi {d_y}}}{\lambda }\sin {\varphi _t}\cos {\eta _t}}}} \right]^\mathrm{T}}\otimes {\left[ {1,{e^{ - j\frac{{2\pi {d_z}}}{\lambda }\sin {\varphi _t}\sin {\eta _t}}}...,{e^{ - j\frac{{2\left( {{M_z} - 1} \right)\pi {d_z}}}{\lambda }\sin {\varphi _t}\sin {\eta _t}}}} \right]^\mathrm{T}}$ \end{scriptsize}\hspace{-0.6em}, \hspace{-0.3em} where $\varphi _t$ and \\ $\eta _t$ are the zenith angle of departure (AoD) and the azimuth AoD of the signal from the STAR-RIS to the user $k$ respectively. $d_{y}$ and $d_{z}$ are the STAR-RIS element spacings along the $y$- and $z$-axis respectively, $\lambda$ denotes the signal wavelength, and $\otimes$ denotes the Kronecker product.

For any user $k$, the downlink equivalent-combined channel gain is given as

\begin{equation}
    {h_{k}} = {h_{B,{k}}} + {\mathbf{h}_{S,{k}}}^\mathrm{H}{\mathbf{V} _{k}}{\mathbf{h}_{B,S}},
\end{equation}

\subsection{Hybrid NOMA signal model}
As mentioned earlier, the two paired users are served simultaneously via NOMA, while $K$ user-pairs are served sequentially via TDMA. Let ${\pi}(k) \in\{0,1\}$ denote the decoding order of user $k$. In downlink NOMA, to guarantee that successive interference cancelation (SIC) performs successfully, an optimal decoding order is to decode in the order of increasing equivalent-combined channel gain, regardless of the user power allocation, that is,
\begin{equation}\label{pi}
\pi \left( {{k}} \right) =  \left\{ \begin{array}{l}
0, if \left| {{h_{{k}}}} \right| \le \left| {{h_{\bar{k}}}} \right|, \\
1, if \left| {{h_{{k}}}} \right| > \left| {{h_{\bar {k}}}} \right|.
\end{array} \right.
\end{equation}
Under the proposed hybrid NOMA framework, the signal-to-interference-plus-noise ratio (SINR) at user $k$ is expressed as
\begin{equation}\label{SINR}
  \Gamma _{{k}} = \frac{{{{\left| {{h_{{k}}}} \right|}^2}{\rho _{{k}}}P}}{{\pi \left( {\bar {k}} \right){{\left| {{h_{{k}}}} \right|}^2} {{\rho_{{\bar {k}}}}}P  + n}},
\end{equation}
where $P$ is the total power of the BS. ${\rho _{k}},\rho _{{\bar{k}}} \in [0,{\rm{1}}]$ are the power allocation coefficients, which satisfy ${\rho _{{k}}}{\rm{ + }}{\rho _{{\bar{k}}}}{\rm{ = 1}}$ . $n$ is the complex circular i.i.d. additive Gaussian noise with $n \sim \mathcal{CN}\left( {0,{\sigma ^2}} \right)$, where ${\sigma ^2}$ is the noise power.
Then, the downlink communication rate of the user is given by
\begin{equation}
    {R_{{k}}} = {\tau _{\{ {k},\overline {k}\} }}{\log _2}\left( {1 + {\Gamma _{{k}}}} \right)
\end{equation}

\subsection{Problem Formulation}
Considering the communication fairness in cell-edge areas, the minimum transmission rate is taken as the performance metric of the proposed STAR-RIS assisted networks. The object is to maximize the minimum transmission rate among multiple users, by jointly optimizing the user pairing, the decoding order, the passive beamforming design, the user-pair time allocation and the user power allocation. The problem is formulated as follows,
\vspace{-3cm}
\begin{subequations}\label{P1}
\begin{align}
\text{(P1):}~~&{\mathop {{\rm{Max}}}\limits_{{\mathbf{V} }_{k},{c},{\pi_{k} },{\tau _{\{ k,\bar{ k}\} }},{\rho _k}}  {\kern 5pt} \mathop {{\rm{Min}}}\limits_{k \in {{\cal K}_L} \cup {{\cal K}_R}} {R_k}}\\
~~\text{s.t.}~~~&{\sqrt {\beta _k} \in [0,1]},{\beta _k  +\beta _{\bar{k}}  = 1}, \label{a}\\
&\theta _k^m \in [0,2\pi ],\label{b}\\
&{{c_{{k_L},{k_R}}} \in \left\{ {0,1} \right\}},{\sum\limits_{{k_R} = 1}^K {{c_{{k_L},{k_R}}} = 1} },\label{c}\\
&\pi \left( k \right) = \left\{ {\begin{array}{*{20}{l}}
{0,if\left| {{h_k}} \right| \le \left| {{h_{\bar k}}} \right|,}\\
{1,if\left| {{h_k}} \right| > \left| {{h_{\bar k}}} \right|,}
\end{array}} \right. \label{d}\\
&{{\tau _{\{ {k},\bar {{k}} \} }} \in [0,1]},\sum\limits_{k \in {{\cal K}_L}({k \in {{\cal K}_R}})} {{\tau _{\{ k,\bar k \} }} = 1},\label{e}\\
&{{\rho _k} \in [0,1],{\rho _k} + {\rho _{\bar k}} = 1}.\label{f}
\end{align}
\end{subequations}(\ref{a}) and (\ref{b}) define the feasible ranges of STAR-RIS amplitude coefficients and phase shift respectively. Constraint (\ref{c}) illustrates the one-to-one mapping relationship between TUs and RUs. Constraint (\ref{d}) guarantees the SIC performs successfully. Constraint (\ref{e}) means the sum of the time allocation coefficients for user-pairs is $1$ and constraint (\ref{f}) represents that the sum of the power allocation coefficients for two paired users should also be $1$.
\vspace{-0.3cm}
\section{Problem Solution and Proposed Algorithm}
Evidently, the proposed optimization problem is nonconvex and highly coupled. Moreover, it is worth noting that the decoding order, beamforming vectors, power and time allocation can only be solved when the user pairing is given, due to the constraints (\ref{a}), (\ref{d}), (\ref{e}) and (\ref{f}). However, when the other variables are given, it is difficult to further optimize the user pairing through mathematical derivation. To tackle this issue, a novel iterative algorithm with a two-layer loop-nesting structure is proposed. Specifically, the outer layer is designed to determine the user pairing by a one-to-one swapping matching based approach. With given user pairing, the decoding order, passive beamforming, user-pair time allocation and user power allocation are solved by an alternating optimization (AO) based algorithm in the inner layer.
\vspace{-0.35cm}
\subsection{Outer layer: one-to-one matching based user-pairing}
In this subsection, we describe the user pairing as a matching game between TUs and RUs, which is solved by a matching-theory based algorithm. First of all, some basic concepts are introduced as follows.
\vspace{-0.15cm}
\begin{Definition}\label{Definition_1}
(One-to-one Two-sides Matching):\emph{ A one-to-one matching $\Phi$ is defined as a function from ${{\cal K}_T} \cup {{\cal K}_R}$ to ${{\cal K}_T} \cup {{\cal K}_R}$ such that
\begin{enumerate}
  \item $\Phi ({k_T}) \in {{\cal K}_R},\Phi ({k_R}) \in {{\cal K}_T},$
  \item $\left|{\Phi ({k_T})}\right|=\left|{\Phi ({k_R})}\right|=1$
  \item ${k_T} = \Phi ({k_R}) \Leftrightarrow {k_R} = \Phi ({k_T})$
\end{enumerate}}
\end{Definition}
\vspace{-0.25cm}
\begin{Definition}\label{Definition_2}
(Utility Function):\emph{ Given a matching state $\Phi$, the utility function ${W}(\Phi )$ is defined as the possible maximum value of the minimum user rate, which is obtained in the subsequent inner layer. }
${W}(\Phi )={\mathop {{\rm{Max}}}\limits_{{\mathbf{V} }_{k},{\pi_{k} },{\tau _{\{ k,\bar{ k}\} }},{\rho _k}}  {\kern 5pt} \mathop {{\rm{Min}}}\limits_{k \in {{\cal K}_L} \cup {{\cal K}_R}} {R_k}}.$
\end{Definition}
\vspace{-0.45cm}
\begin{Definition}\label{Definition_3}
(Swapping Matching):
\emph{For matching state $\Phi ({k_L}) = {k_R}$ and $\Phi ({{\tilde k}_L}) = {{\tilde k}_R}$, a swap matching is
\begin{small}
$\Phi _{{k_L}}^{{{\tilde k}_L}} = \left\{ \Phi \backslash \left\{ ({k_L},{k_R}),\left({{\tilde k}_L},{{\tilde k}_R}\right) \right\} \cup \left\{ \left({k_L},{{\tilde k}_R}\right),\left({{\tilde k}_L},{k_R}\right) \right\} \right\}$
\end{small}.}
\end{Definition}
\begin{Definition}\label{Definition_4}
(Swap-blocking pair):\emph{ Given a matching state $\Phi$ with $ \small{\Phi \left({k_T}\right) = {k_R}}$ and $\small{\Phi \left({{\tilde k}_T}\right) = {{\tilde k}_R}}$, $\small{\left({k_T}, {{\tilde k}_T}\right)}$ is a swapping-blocking pair if and only if $\small{{W}\left(\Phi _{{k_T}}^{{{\tilde k}_T}}\right) > {W}\left(\Phi \right)}$.}
\end{Definition}
\vspace{-0.2cm}

Definition \ref{Definition_1} indicates the one-to-one mapping relationship between TUs and RUs. Note that different from general matching algorithms, the utility function in Definition \ref{Definition_2} is a global function, which makes the obtained matching state closer to the optimal solution, rather than just a stable matching. Definition \ref{Definition_3} enables two TUs to exchange their paired RUs. Definition \ref{Definition_4} implies that the utility function will increase with a swap-blocking pair. Based on the above definitions, the one-to-one swapping matching based user pairing in the outer layer is briefly described as: with a given initial matching state, the algorithm keeps searching for one TU and one RU to form a swapping-blocking pair and executes the swapping matching until there is no swapping-blocking pair.
\vspace{-0.2cm}
\subsection{Inner layer: AO based decoding order, passive beamforming, power and time allocation}
In this subsection, we aim to obtain the utility function defined in the outer layer. With given user pairing, the problem for jointly optimizing the decoding order, the beamforming, the power and time allocation is expressed as
\begin{subequations}\label{P2}
\begin{align}
\vspace{-0.3cm}
\text{(P2):}~~&{\mathop {{\rm{Max}}}\limits_{S,{{\bf{V}}_k},{\pi _k},{\tau _{\{ k,\bar k\} }},{\rho _k}} S}\\
~~\text{s.t.}~~~&{S \le {R_k},k \in {\mathcal{K}_T} \cup {\mathcal{K}_R}}, (\ref{a}),(\ref{b}),(\ref{d})\sim(\ref{f}), \label{P2a}
\vspace{-0.3cm}
\end{align}
\end{subequations}where $S$ is an introduced auxiliary variable which represents the minimum rate among users. To seek the solution of the still highly coupled nonconvex problem (P1), an AO based iteration algorithm is proposed, where one variable is solved with giving the other ones \footnote{With sufficient iterations, the order in which multiple variables are solved has little effect on the final results.}.

First of all, as considered in previous works \cite{STAR+NOMA03,decoding_order2}, the decoding order can be determined by the equivalent-combined channel gain when the other variables are given. Furthermore,
the SIC constraint can be removed under the optimal decoding order \begin{small}$\pi^\ast \left( k \right) = \left\{ {\begin{array}{*{20}{l}}
{0,if\left| {{h_k}} \right| \le \left| {{h_{\bar k}}} \right|,}\\
{1,if\left| {{h_k}} \right| > \left| {{h_{\bar k}}} \right|.}
\end{array}} \right.$ \end{small}This operation will not affect the solution of (P2) by iteratively updating the optimal decoding order and solving the problem (P2) without (\ref{d}).

Then, we focus on the passive beamforming design at the STAR-RIS. It is known that the beamforming is expected to maximize the equivalent-combined channel gain. Based on the triangle inequality, there exists
$\small{\left| {{h_k}} \right| = \left| {{h_{B,k}} + {\mathbf{h}_{S,k}}^\mathrm{H}{\mathbf{V}_k}{\mathbf{h}_{B,S}}} \right|\mathop  \le \limits^{(a)} \left| {{h_{B,k}}} \right| + \left| {{\mathbf{h}_{S,k}}^\mathrm{H}{\mathbf{V}_k}{\mathbf{h}_{B,S}}} \right|}$, where (a) holds if and only if $\small{\arg ({h_{B,k}}) = \arg ({\mathbf{h}_{S,k}}^\mathrm{H}{\mathbf{V}_k}{\mathbf{h}_{B,S}}) = \varphi _k^0}$. Actually, this indicates that the optimal beamforming is supposed to align the signals from the direct link and that from the STAR-RIS assisted link. Let $\small{{\mathbf{\Theta} _k} = {\left[ {{e^{j\theta _k^1}},{e^{j\theta _k^2}},...,{e^{j\theta _k^M}}} \right]^\mathrm{H}}}$ and
$\small{{\mathbf{e}_k} = diag({\mathbf{h}_{S,k}}^\mathrm{H}){\mathbf{h}_{B,S}}}$, it is easy to know $\small{{\mathbf{h}_{S,k}}^\mathrm{H}{\mathbf{V}_k}{\mathbf{h}_{B,S}} = \sqrt {{\beta _k}} {\mathbf{\Theta} _k}^\mathrm{H}{\mathbf{e}_k}}$. As a result, the optimal phase shift at the STAR-RIS is given as
\begin{equation}\label{phase}
\small{
\hspace{-0.35em} \theta _k^m \hspace{-0.2em}= \hspace{-0.2em} \bmod \hspace{-0.2em} \left[ {\arg ({h_{B,k}}) \hspace{-0.2em}- \hspace{-0.2em}\arg ({\mathbf{h}_{S,k}}^\mathrm{H}[m])\hspace{-0.2em} - \hspace{-0.2em}\arg ({\mathbf{h}_{B,S}}[m]),2\pi } \right].}
\end{equation}

As for the beamforming amplitude ${\beta _k}$, to tackle the nonconvex constraint $S \le {R_k}$, a successive convex approximation (SCA) based approach is applied. Since the signals from two links are aligned, the channel gain can be rewritten as
\begin{equation}
{\left| {{h_k}} \right|^2} = A_k+B_k\sqrt {{\beta _k}} +C_k{\beta _k},
\end{equation}
where $A_k={\left| {{h_{B,k}}} \right|^2}$, $B_k=2\left| {{h_{B,k}}} \right|\left| {{\mathbf{h}_{S,k}}^\mathrm{H}diag({\mathbf{\Theta} _k}){\mathbf{h}_{B,S}}} \right|$ and $C_k={\left| {{\mathbf{h}_{S,k}}^\mathrm{H}diag({\mathbf{\Theta} _k}){\mathbf{h}_{B,S}}} \right|^2}$. With ${L_{k,1}} = P{\rho _k}$ and ${L_{k,2}} = \pi (\bar k)P{\rho _{\bar k}}$, the constraint $S \le {R_k}$ is transformed as
\begin{equation}\label{belta}
\small{
\begin{split}
S+{\tau _{\{ k,\bar k\} }}{\log _2}\left( {{L_{k,2}}\left( {A_k + B_k\sqrt {{\beta _k}}  + C_k{\beta _k}} \right) + n} \right)\le\\
 {\tau _{\{ k,\bar k\} }} {\log _2}\left( {\left( {{L_{k,1}} + {L_{k,2}}} \right)\left( {A_k + B_k\sqrt {{\beta _k}}  + C_k{\beta _k}} \right) + n} \right).
\end{split}}
\end{equation}
As ${\log _2}\left( {\sqrt x  + x} \right)$ is a concave function with regard to $x$, we consider taking the
first-order Taylor expansion at the given feasible points ${\beta _k}^\varepsilon $ to obtain the upper bound of ${\log _2}\left( {{L_{k,2}}\left( {A_k + B_k\sqrt {{\beta _k}}  + C_k{\beta _k}} \right) + n} \right))$:
\begin{small}
\begin{equation}
\begin{split}
  \Xi _{k,1}^{UB} = {\log _2}\left( {{L_{k,2}}\left( {{A_k} + {B_k}\sqrt {{\beta _k}^\varepsilon }  + {C_k}{\beta _k}^\varepsilon } \right) + n} \right){\rm{ + }}\\
   \frac{{{L_{k,2}}\left( {{{{B_k}} \mathord{\left/{\vphantom {{{B_k}} {{\rm{2}}\sqrt {{\beta _k}^\varepsilon } }}} \right.
\kern-\nulldelimiterspace} {{\rm{2}}\sqrt {{\beta _k}^\varepsilon } }}{\rm{  +  }}{{\rm{C}}_k}} \right)}}{{{\rm{log(2)}}\left( {{\rm{n  +  }}{L_{k,2}}\left( {{A_k} + {B_k}\sqrt {{\beta _k}^\varepsilon }  + {C_k}{\beta _k}^\varepsilon } \right)} \right)}}({\beta _k} - {\beta _k}^\varepsilon ).
\end{split}
\end{equation}
\end{small}To make (\ref{belta}) more tractable, auxiliary variables ${\xi _k} \le A + B\sqrt {{\beta _k}}  + C{\beta _k}$  are introduced and the subproblem to optimize the beamforming amplitude is converted into
\begin{subequations}\label{P5}
\small{
\begin{align}
\hspace{-0.6em}\text{(P2-1):}~~&\mathop {{\rm{Max}}}\limits_{\beta _k,\xi _k} S,\\
~~\text{s.t.}~~&S \hspace{-0.1em}+\hspace{-0.1em} {\tau _{\{ k,\bar k\} }}\Xi _{k,1}^{UB} \hspace{-0.1em}\le \hspace{-0.1em}{\tau _{\{ k,\bar k\} }}{\log _2}\left( {\left( {{L_{k,1}}\hspace{-0.1em} +\hspace{-0.1em} {L_{k,2}}} \right){\xi _k} \hspace{-0.1em}+ \hspace{-0.1em}n} \right),\label{P5a}\\
&{\xi _k} \le A_k + B_k\sqrt {{\beta _k}}  + C_k{\beta _k},\label{P5b}\\
&(\ref{a}),
\end{align}}
\end{subequations}
which can be solved by CVX.

Next, given $\left\{ {{\pi _k},{\mathbf{V}_k},{\tau _{\{ k,\bar k\} }}} \right\}$, $S \le {R_k}$  with regard to the user power allocation can be rewritten as
\begin{equation}
\small{
  \hspace{-0.9em} S \hspace{-0.2em}+\hspace{-0.2em} {\tau _{\{ k,\bar k\} }}{\log _2}\left( {{L_{k,3}}{\rho _{\bar k}}\hspace{-0.2em} +\hspace{-0.2em} n} \right) \hspace{-0.2em}\le \hspace{-0.2em}{\tau _{\{ k,\bar k\} }}{\log _2}\left( {{L_{k,4}}{\rho _k} \hspace{-0.2em}+ \hspace{-0.2em}{L_{k,3}}{\rho _{\bar k}}\hspace{-0.2em}+\hspace{-0.2em}n}\hspace{-0.1em}\right),}
\end{equation}
where $\small{{L_{k,3}} = \pi (\bar k)P{\left| {{h_k}} \right|^2}}$ and $\small{{L_{k,4}} = P{\left| {{h_k}} \right|^2}}$. Similar to (\ref{belta}), the upper bound of $\small{{\log _2}\left( {{L_{k,3}}{\rho _{\bar k}} + n} \right)}$ can be derived by the first-order Taylor expansion at the given feasible points ${\rho _k}^\varepsilon $ and ${\rho _{\bar k}}^\varepsilon $:
\begin{equation}
\small{
\hspace{-0.5em} \Xi _{k,2}^{UB} \hspace{-0.1em} = \hspace{-0.1em}{\log _2}\left( {{L_{k,3}}{\rho _{\bar k}}^\varepsilon  + n} \right)\hspace{-0.1em} + \hspace{-0.1em}\frac{{{L_{k,3}}}}{{\log 2\left( {{L_{k,3}}{\rho _k}^\varepsilon \hspace{-0.1em} +\hspace{-0.1em} n} \right)}}\left({\rho _{\bar k}}\hspace{-0.1em} - \hspace{-0.1em}{\rho _{\bar k}}^\varepsilon\right). \hspace{-0.1em}}
\end{equation}
As a result, the subproblem to solve the user power allocation is expressed as
\begin{small}
\begin{subequations}
\begin{align}
\text{(P2-2):}~~&\mathop {{\rm{Max}}}\limits_{\rho  _k} S,\\
~~\text{s.t.}~~~&S+\Xi _{k,2}^{UB} \le {\tau _{\{ k,\bar k\} }}{\log _2}\left( {{L_{k,4}}{\rho _k} + {L_{k,3}}{\rho _{\bar k}} + n} \right),\\
&(\ref{f}).
\end{align}
\end{subequations}
\end{small}

Finally, the subproblem to solve the user-pair time allocation can be expressed as a linear programming problem:
\begin{subequations}
\small{
\begin{align}
\text{(P2-3):}~~&\mathop {{\rm{Max}}}\limits_{{\tau _{\{ k,\bar k\} }}} S,\\ \vspace{-0.7cm}
\text{s.t.}~~~&S \le {\tau _{\{ k,\bar k\} }}{log_2}\left(1+{\frac{{{{\left| {{h_{{k}}}} \right|}^2}{\rho _{{k}}}P}}{{\pi \left( {\bar {k}} \right){{\left| {{h_{{k}}}} \right|}^2} {{\rho _{{\bar {k}}}}}P  + n}}}\right),\\ \vspace{-0.7cm}
&(\ref{e}).
\end{align}}
\end{subequations}
 \vspace{-1.0cm}
\subsection{Discussion on the Proposed Algorithm}
In summary, the proposed iterative algorithm has a two-layer loop-nesting structure. The details are shown in Algorithm \ref{A1}.
\vspace{-0.5cm}
\begin{algorithm}[hptb]
\begin{small}
	\caption{The proposed two-layer iterative algorithm}
	\label{A1}
\begin{algorithmic}[1]
    \vspace{-0.05cm}
	\STATE \textbf{Initialization:} user pairing $c$. \vspace{-0.04cm}
    \STATE \textbf{Repeat}\vspace{-0.04cm}
    \STATE \hspace{6pt}\textbf{Initialization:} ${\mathbf{V} }_{k}^0$, $\rho _k^0$ and $\tau _{\{ k,\bar k\} }^0$,\vspace{-0.04cm}
    \STATE \hspace{6pt}\textbf{Repeat} \vspace{-0.04cm}
    \STATE \hspace{12pt}$\varepsilon=1$, \vspace{-0.04cm}
    \STATE \hspace{12pt}Calculate the equivalent-combined channel gain and determine the optimal decoding order ${\pi _k}^\ast$ according to (\ref{pi}), \vspace{-0.04cm}
    \STATE \hspace{12pt}Obtain the optimal phase shift $\theta {_k^\ast}$ according to (\ref{phase}), \vspace{-0.04cm}
    \STATE \hspace{12pt}Obtain the optimal amplitude ${\beta _k}^\ast$ by solving (P2-1), \vspace{-0.04cm}
    \STATE \hspace{12pt}Obtain the optimal power allocation ${\rho _k}^\ast$ by solving (P2-2), \vspace{-0.04cm}
    \STATE \hspace{12pt}Obtain the optimal time allocation $\tau {_{\{ k,\bar k\} }}^\ast$ by solving (P2-3), \vspace{-0.25cm}
    \STATE \hspace{12pt}Update $\small{\left\{ {{V_k}^\varepsilon ,{\rho _k}^\varepsilon ,\tau _{\{ k,\bar k\} }^\varepsilon } \right\} \leftarrow \left\{ {{V_k}^\ast ,{\rho _k}^\ast ,\tau _{\{ k,\bar k\} }^\ast } \right\}}$, \vspace{-0.04cm}
    \STATE \hspace{12pt}Update $\varepsilon  \leftarrow \varepsilon  + 1$, \vspace{-0.04cm}
    \STATE \hspace{6pt}\textbf{Until} the inner layer algorithm reaches the accuracy or inner iterates beyond the permitted maximum times. \vspace{-0.045cm}
    \STATE \hspace{6pt}Obtain the utility function by the inner layer algorithm. \vspace{-0.045cm}
    \STATE \hspace{6pt}For any TU ${k_T}$, it search for all other TUs ${\tilde k}_T$ \vspace{-0.04cm}
    \STATE \hspace{6pt}\textbf{if} $\small{\left( {{k_T},{{\tilde k}_T}} \right)}$ is a swap-blocking pair \vspace{-0.045cm}
    \STATE \hspace{12pt} Update $\Phi  = \Phi _{{k_T}}^{{{\tilde k}_T}}$, \vspace{-0.04cm}
    \STATE \hspace{6pt}\textbf{end if} \vspace{-0.045cm}
    \STATE \textbf{Until} the outer layer algorithm reaches the accuracy or outer iterates beyond the permitted maximum times. \vspace{-0.045cm}
\end{algorithmic}
\end{small}
\end{algorithm}
\vspace{-0.3cm}

1) Convergence: For the outer layer, the objective value is non-decreasing after each iteration due to the definition of the swap-blocking pair. For the alternating iteration in the inner layer, the objective value is also non-decreasing, which is proven as follows.
Firstly, in steps 6, 7 and 10 of Algorithm \ref{A1}, since the optimal solution can be obtained by (\ref{pi}), (\ref{phase}) and solving (P2-3), we have $\footnotesize{S\left(\pi _k^\varepsilon ,\theta _k^\varepsilon ,\beta _k^\varepsilon ,\rho _k^\varepsilon ,\tau _{\{ k,\bar k\} }^\varepsilon \right) \le S\left(\pi _k^{\varepsilon  + 1},\theta _k^{\varepsilon  + 1},\beta _k^\varepsilon ,\rho _k^\varepsilon ,\tau _{\{ k,\bar k\} }^{\varepsilon  + 1}\right)}$.
Secondly, define $S_{amp}^{lb,\varepsilon}$ as the objective value obtained by solving (P2-1) at the $\varepsilon$-th iteration. For step 8, it follows
\vspace{-0.2cm}
\begin{footnotesize}
\begin{align*}
S\left(\hspace{-0.2em}\pi _k^{\varepsilon  + 1},\hspace{-0.2em} \theta _k^{\varepsilon  + 1},\hspace{-0.2em} \beta _k^\varepsilon ,\hspace{-0.2em} \rho _k^\varepsilon,\hspace{-0.2em} \tau _{\{ k,\bar k\} }^{\varepsilon  + 1}\right) \hspace{-0.2em}
& \mathop  = \limits^{(a)}\hspace{-0.2em} S_{amp}^{lb,\varepsilon }\left(\pi _k^{\varepsilon  + 1},\theta _k^{\varepsilon  + 1},\beta _k^\varepsilon ,\xi _k^\varepsilon , \rho _k^\varepsilon ,\tau _{\{ k,\bar k\} }^{\varepsilon  + 1}\right) \\ \vspace{-0.2cm}
\vspace{-0.1cm}
& \mathop \leq \limits^{(b)}\hspace{-0.2em} S_{amp}^{lb,\varepsilon }\hspace{-0.2em}\left(\hspace{-0.2em}\pi _k^{\varepsilon  + 1},\hspace{-0.2em} \theta _k^{\varepsilon  + 1},\hspace{-0.2em} \beta _k^{\varepsilon  + 1},\hspace{-0.2em}  \xi _k^{\varepsilon+1} ,\hspace{-0.2em} \rho _k^\varepsilon,\hspace{-0.2em} \tau _{\{ k,\bar k\} }^{\varepsilon  + 1}\right) \\ \vspace{-0.2cm}
&\mathop \leq \limits^{(c)}\hspace{-0.2em} S\left(\pi _k^{\varepsilon  + 1},\theta _k^{\varepsilon  + 1},\beta _k^{\varepsilon  + 1},\rho _k^\varepsilon ,\tau _{\{ k,\bar k\} }^{\varepsilon  + 1}\right), \vspace{-0.2cm}
\end{align*}
\end{footnotesize}where (a) holds since the first-order Taylor expansion in (13) is tight at the given local points; (b) holds because of the optimized $\beta _k^{\varepsilon  + 1}$ and $\xi _k^{\varepsilon+1} $; (c) holds since the objective value of (P2-1) is the lower bound of that of the original problem (P2) at $\beta _k^{\varepsilon  + 1}$. Similarly, for step 9, there is \begin{footnotesize} $\hspace{-0.2em}S\left(\hspace{-0.2em}\pi _k^{\varepsilon+ 1},\hspace{-0.2em}\theta _k^{\varepsilon+ 1},\hspace{-0.2em}\beta _k^{\varepsilon+ 1},\hspace{-0.2em}\rho _k^\varepsilon,\hspace{-0.2em}\tau _{\{ k,\bar k\} }^{\varepsilon+ 1}\right) \hspace{-0.2em}\le\hspace{-0.2em} S\hspace{-0.2em}\left(\pi _k^{\varepsilon + 1},\hspace{-0.2em}\theta _k^{\varepsilon+ 1},\hspace{-0.2em}\beta _k^{\varepsilon+ 1},\hspace{-0.2em}\rho _k^{\varepsilon+ 1},\hspace{-0.2em}\tau _{\{ k,\bar k\} }^{\varepsilon+ 1}\right)$. \end{footnotesize}
Based on the above analysis, it can be driven that \begin{footnotesize} $S\left(\pi _k^\varepsilon ,\theta _k^\varepsilon ,\beta _k^\varepsilon ,\rho _k^\varepsilon ,\tau _{\{ k,\bar k\} }^\varepsilon \right) \le S\left(\pi _k^{\varepsilon  + 1},\theta _k^{\varepsilon  + 1},\beta _k^{\varepsilon+1} ,\rho _k^{\varepsilon+1} ,\tau _{\{ k,\bar k\} }^{\varepsilon+ 1}\right)$, \end{footnotesize} and the proof is completed.
Since the achievable max-min rate is upper bounded by a finite value, the proposed algorithm is guaranteed to converge.

2) Complexity: The computational complexity of Algorithm \ref{A1} mainly depends on the number of users. The maximum number of swap operations in the outer layer is $K\left( {K - 1} \right)/2$. The complexity for solving (P2) in the inner layer is \begin{small}
${\cal O}\left( {{I_{{\rm{ite}}}}\left( {2{{\left( {K} \right)}^{4.5}} + {K^{2}}(K+1)} \right)} \right)$
\end{small}
if the interior point method is employed, where ${{I_{{\rm{ite}}}}}$ denotes the iterations taken to converge. Then the overall computational complexity of Algorithm \ref{A1}  is
\begin{small}
$O\left( {K\left( {K - 1} \right){{I_{{\rm{ite}}}}\left( {2{{\left( {K} \right)}^{4.5}} + {K^{2}}(K+1)} \right)}/2} \right)$, i.e.,$O\left( {{I_{{\rm{ite}}}}{K^{{\rm{6}}{\rm{.5}}}}} \right)$.
\end{small}

\section{ Numerical Results}

In this section, numerical results are provided to demonstrate the effectiveness of the proposed framework and algorithm. We consider an area with the size of $1000$m $\times$ $1000$m, where the STAR-RIS is deployed at the center. The main simulation setups are shown in  Table \ref{table1}.
\begin{table}[htbp]
	\caption{Simulation Parameters}
    \label{table1}
	\begin{center}
		\begin{tabular}{|p{4.1cm}<{\centering}|c|}
            \hline
            \textbf{Parameter} & \textbf{Value}  \\       \hline
	        Communication bandwidth            & $BW = 1$ MHz \\ \hline
            BS transmission power   & ${P_{\max}} = 30$ dBm \\ \hline
            UAV fight height                 & ${H_U} = 200$ m \\ \hline
            Carrier frequency               & ${f_c} = 750$ MHz \\ \hline
            Wavelength                      & $\lambda = 3\times10^{8}/{f_c}$ \\ \hline
            STAR-RIS element spacings     & $d_{y}=d_{z}= \lambda/10$ \\ \hline
            Noise power dense                    & ${\sigma ^2} = -150$ dBm/Hz  \\ \hline
            Path loss exponents   & ${\alpha _1} = 3$, ${\alpha _2} = 2$, ${\alpha _3} = 2.3$\\ \hline
		\end{tabular}
	\end{center}
    \vspace{-0.6cm}
\end{table}

\begin{figure}[htbp]
        \centering
		\includegraphics[width=2.8in]{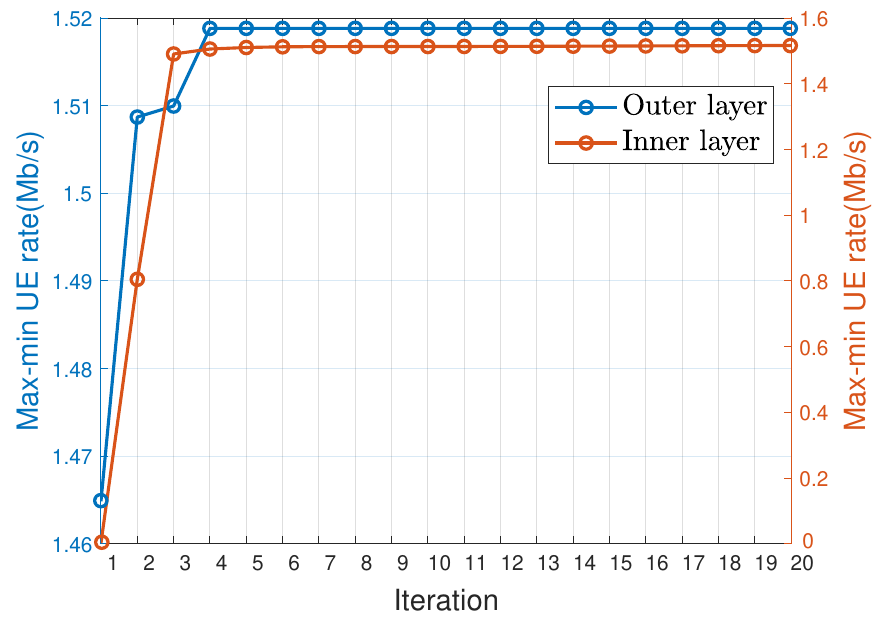}\\
		 \caption{Convergence of the proposed algorithm.}\label{Fig2}
\end{figure}

\begin{figure*}[htbp]
	\centering
	\begin{minipage}{0.329\linewidth}
		\centering
		\includegraphics[width=2.5in]{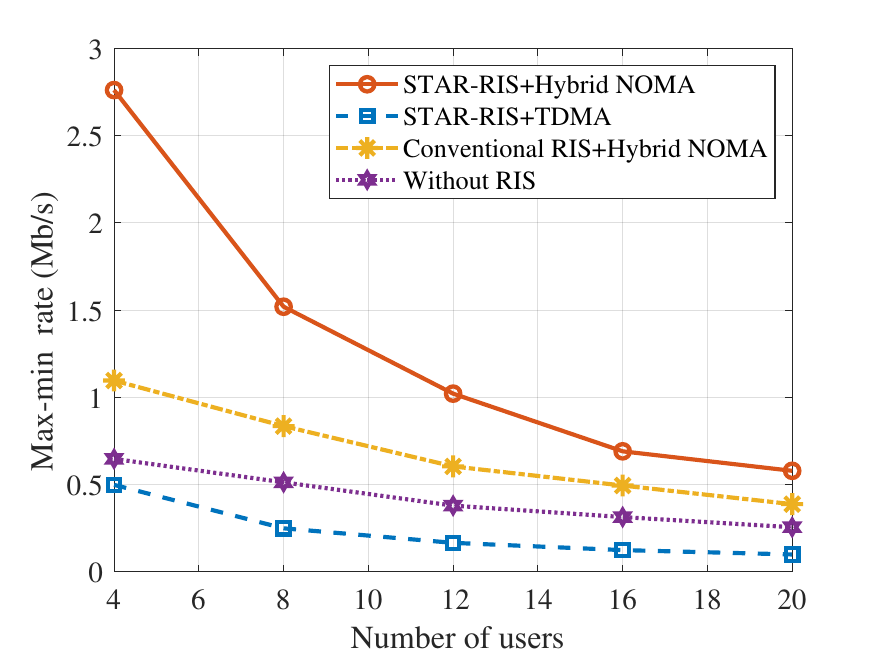}\\
		 \caption{Max-min rate with different frameworks vs. the number of users}\label{Fig3}
   	\end{minipage}
	\begin{minipage}{0.329\linewidth}
		\centering
		\includegraphics[width=2.5in]{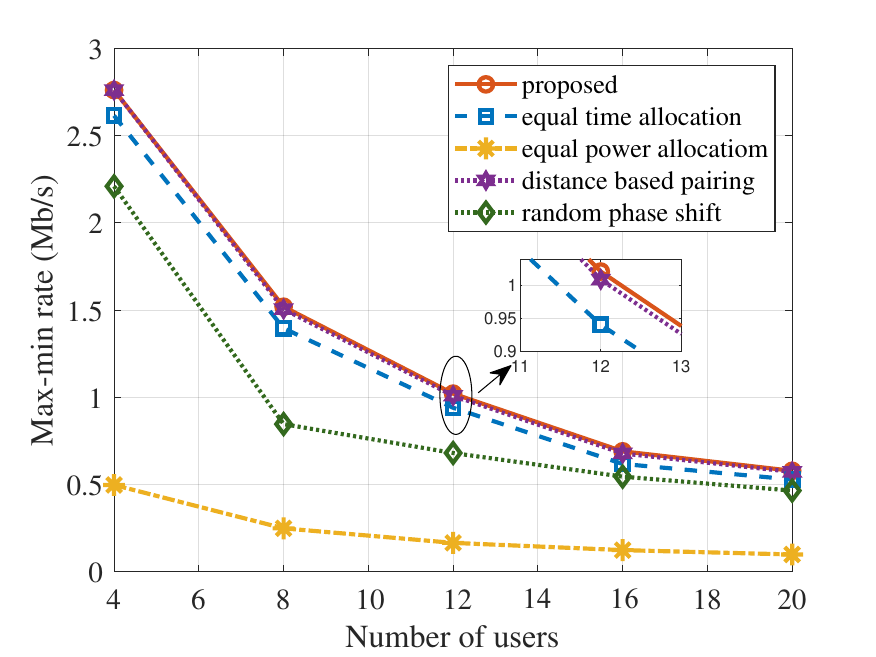}\\
        \caption{Max-min rate with different algorithms vs. the number of users}\label{Fig4}
	\end{minipage}
    \begin{minipage}{0.329\linewidth}
		\centering
		\includegraphics[width=2.5in]{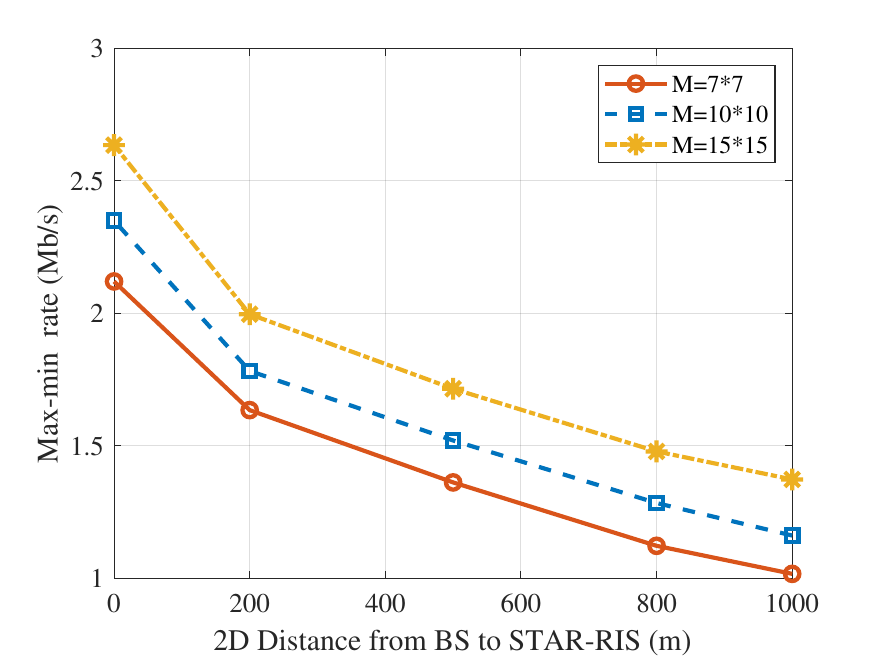}\\
        \caption{Max-min rate with different $M$ vs. the 2D distance from BS to STAR-RIS}\label{Fig5}
	\end{minipage}
\end{figure*}

Fig. \ref{Fig2} illustrates the convergence process of the proposed two-layer iterative algorithm. We can see that both inner-layer iteration and outer-layer iteration show a growing trend and achieve convergence within 10 iterations, which verifies the feasibility of the proposed algorithm. In addition, it is shown that the performance gain obtained by the outer-layer iteration is significantly smaller than that obtained from the inner-layer iteration, which reveals that the user pairing optimization in the outer layer has a modest impact on network performance. This issue will be further demonstrated in Fig. \ref{Fig4}.

In Fig. \ref{Fig3}, we evaluate the performance of the proposed hybrid NOMA based STAR-RIS assisted network. There are three benchmark frameworks including TDMA based STAR-RIS assisted network,  hybrid NOMA based reflecting-only RIS assisted network, and hybrid NOMA based network without RIS. It can be observed that regardless of the frameworks, the max-min user rate decreases as the number of users increases. This is caused by the wireless resource competition among multiple users. Compared with the three benchmarks, we can find that the combination of STAR-RIS and hybrid NOMA has the best performance. Specifically, the gain from the hybrid NOMA is greater than that from the STAR-RIS, with 6.08 times and 2.96 times respectively when $K = 8$. The max-min user rate of the proposed framework is also significantly improved compared with that of conventional reflecting-only RIS assisted network. This improvement exactly comes from the fact that the  STAR-RIS can transmit and reflect signals simultaneously, allowing it to serve all users, regardless of its location.

In Fig. \ref{Fig4}, the performance of the proposed two-layer iterative algorithm is compared with following benchmark algorithms: 1) equal user-pair time allocation, ${\tau _{\{ k,\bar k\} }} = 1/K$; 2) equal user power allocation, ${\rho _k} = {\rho _{\bar k}} = 0.5$; 3) distance based user pairing, which means the matching state is determined by the distance from users to the STAR-RIS (the far TU is paired with the near RU); and 4) random phase shift at the STAR-RIS. As we can see, the max-min user rate of the proposed algorithm is the largest among all five algorithms, which verifies the superiority of the proposed algorithm. Furthermore, the gain from power optimization is the largest, followed by that from beamforming design, while the gains from other variables are relatively small. This result shows the critical importance of power allocation for NOMA and beamforming design for STAR-RIS.

Both Fig. \ref{Fig3} and Fig. \ref{Fig4} are obtained with ${M_y} = {M_z} = 10$ and ${\mathbf{q}_S} = [{0},{0},{20}]$ m, while Fig. \ref{Fig5} further investigates the impact of the number of STAR-RIS elements and the location of the STAR-RIS on the network performance \footnote{ The location of the STAR-RIS is considered in a simple way, that is, the distance to the BS. More accurate STAR-RIS deployment will be studied in our future works.}. Based on the results shown in Fig. \ref{Fig5}, two major conclusions can be drawn. On the one hand, increasing the number of elements can improve the enhancement effect of the STAR-RIS. On the other hand, as the max-min rate increases with the shortening of the distance from BS to STAR-RIS, the deployment of STAR-RIS is of great importance.

\section{Conclusion}
In this paper, we investigated a STAR-RIS enhanced cell-edge network, where a novel hybrid NOMA framework was proposed to take full advantages of the STAR-RIS. A max-min user rate problem was solved by jointly optimizing the user pairing, decoding order, passive beamforming, power and time allocation. Numerical results verified the significant superiority of the proposed framework in improving communication fairness. The importance of power allocation and passive beamforming design was also proved. Moreover, it was confirmed that increasing the STAR-RIS elements and shortening the distance of BS and STAR-RIS contributed to the improvement of network performance.

\bibliographystyle{IEEEtran}
\bibliography{mybib}

%

\end{document}